\documentclass[preprint,showpacs,preprintnumbers,amsmath,amssymb]{revtex4}
\usepackage{amsmath}

\usepackage{graphicx}
\usepackage{pdfpages}
\usepackage{dcolumn}
\usepackage{bm}
\usepackage{amsmath} 
\usepackage{amsfonts}
\usepackage{amssymb}
\usepackage{url}
\usepackage{microtype}
\usepackage{graphicx}
\usepackage{appendix}%

\newcommand{\qed}{\nobreak \ifvmode \relax \else
      \ifdim\lastskip<1.5em \hskip-\lastskip
      \hskip1.5em plus0em minus0.5em \fi \nobreak
      \vrule height0.75em width0.5em depth0.25em\fi}

\begin{document}

\preprint{}
\title{Rational-Valued, Small-Prime-Based Qubit-Qutrit and Rebit-Retrit Rank-4/Rank-6 Conjectured  Hilbert-Schmidt Separability Probability Ratios}
\author{Paul B. Slater}
 \email{slater@kitp.ucsb.edu}
\affiliation{%
Kavli Institute for Theoretical Physics, University of California, Santa Barbara, CA 93106-4030\\
}
\date{\today}
            
\begin{abstract}
We implement a procedure--based on the Wishart-Laguerre distribution--recently outlined by both K. {\.Z}yczkowski and the group of A. Khvedelidze, I. Rogojin and V. Abgaryan, for the generation of random (complex or real) $N \times N$ density matrices of rank $k \leq N$ with respect to  Hilbert-Schmidt (HS) measure. In the complex case, one commences with a Ginibre matrix (of normal variates) $A$ of dimensions $k \times k+ 2 (N-k)$, while for a real scenario, one employs a Ginibre matrix $B$ of dimensions $k \times k+1+ 2 (N-k)$. Then, the  $k \times k$ product $A A^{\dagger}$ or $B B^T$ is diagonalized--padded with 
zeros to size $N \times N$--and rotated by a random unitary or orthogonal matrix to obtain a random density matrix with respect to HS measure. Implementing the indicated procedure for rank-4 rebit-retrit states, for 800 million Ginibre-matrix realizations, 6,192,047 were found separable, for a sample probability of .00774006--suggestive of an exact value of $\frac{387}{5000} =\frac{3^2 \cdot 43}{2^3 \cdot 5^4}=.0774$. A prior conjecture for the HS separability probability of rebit-retrit systems of full rank is 
$\frac{860}{6561} =\frac{2^2 \cdot 5 \cdot 43}{3^8} \approx 0.1310775$ (while the two-rebit counterpart has been proven to 
be $\frac{29}{64}=\frac{29}{2^6}$, and the two-qubit one, very strongly indicated to be $\frac{8}{33}=\frac{2^3}{3 \cdot 11}$). Subject to these two conjectures, the ratio of the rank-4 to rank-6 probabilities would be 
$\frac{59049}{1000000}=\frac{3^{10}}{2^6 \cdot 5^6} \approx 0.059049$, with the common factor 43 cancelling. As to the intermediate rank-5 probability, application of a 2006 theorem of Szarek, Bengtsson and {\.Z}ycskowski informs us that it must be one-half the rank-6 probability--itself conjectured to be $\frac{27}{1000} =\frac{3^3}{2^3 \cdot 5^3}$, while for rank 3 or less, the associated probabilities must be 0 by a 2009 result of Ruskai and Werner. We are led to re-examine a 2005 qubit-qutrit analysis of ours, in these regards, and now find evidence for a $\frac{70}{2673}=\frac{2 \cdot 5 \cdot 7}{ 3^5  \cdot 11} \approx 0.0261878$ rank-4 to rank-6 probability ratio.

\end{abstract}
 

\maketitle

\section{Introduction}
Pursuit of the problem ``of quantum separability or inseparability from a measurement theoretical point of view" posed in 1998 by {\.Z}yczkowski, Horodecki, Sanpera and Lewenstein \cite{zyczkowski1998volume} has generated a considerable literature \cite{slater1999priori,slater1999bures,slater2000exact,slater2005qubit,slater2005silver,szarek2006structure,slater2007dyson,slater2013concise,milz2014volumes,fei2016numerical,lovas2017invariance,khvedelidze2017generation,slater2018master,slater2019numerical}.

Of particular interest is the finding--motivated by the results reported in \cite{slater2005qubit}--that the Hilbert-Schmidt PPT (positive-partial-transpose) probability of the generic class of $N \times N$ density matrices of rank $N-1$ is one-half the probability of the density matrices of full rank ($N$) \cite{szarek2006structure}. (For $N = 4, 6$, PPT-probability is equivalent to separability probability.) However, the interesting line of geometric reasoning (Archimedes' formula,\ldots) applied in \cite{szarek2006structure} does not seem extendable to density matrices of rank $k=N-n$, for $n>1$, so further investigative approaches seem necessary in such regards.

In \cite{khvedelidze2017generation}, questions of this (reduced rank) nature were posed. However, the Hilbert-Schmidt separability probability of 0.1652 reported for the rank-three two-qubit states seemed inconsistent with the indicated analysis of Szarek, Bengtsson and {\.Z}yzckowski \cite{szarek2006structure}, since the evidence (both numerical and analytical)--though yet short of a formalized proof--is highly compelling that the Hilbert-Schmidt separability probability of full-rank (4) two-qubit  states is $\frac{8}{33} \approx 0.242424$ \cite{fei2016numerical,milz2014volumes,slater2018master}.

We now implement a procedure--based on the Wishart-Laguerre distribution \cite{livan2018classical,abgaryanprobability}--recently outlined in email communications by both K. {\.Z}yczkowski and the group of A. Khvedelidze, I. Rogojin and V. Abgaryan for the generation of random (complex or real) $N \times N$ density matrices of rank $k$ with respect to  Hilbert-Schmidt measure. In the complex case, one commences with a Ginibre matrix (of normal variates) $A$ of dimensions $k \times k+ 2 (N-k)$, while for a real scenario, one employs a Ginibre matrix $B$ of dimensions $k \times k+1+ 2 (N-k)$. Then, the $k \times k$ product $A A^{\dagger}$ or $B B^T$ is diagonalized, padded with zeros to size $N\times N$, and then rotated by a random unitary or orthogonal matrix to obtain, as desired, a random density matrix with respect to Hilbert-Schmidt measure.

In \cite{slater2019numerical}, conjectures of Hilbert-Schmidt separability probabilities of $\frac{860}{6561} =\frac{2^2 \cdot 5 \cdot 43}{3^8} \approx 0.1310775$ and $\frac{27}{1000}= \frac{3^3}{2^3 \cdot 5^3} \approx 0.027$ were advanced--based on 1,850,000,000 and 2,415,000,000 iterations--for generic rebit-retrit and qubit-qutrit systems, respectively. (In \cite[Tab. 1]{khvedelidze2017generation} a qubit-qutrit probability estimate of 0.0270 was reported.) Additionally, in the 2005 study \cite{slater2005qubit}, the rank-4 qubit-qutrit Hilbert-Schmidt separability probability was reported to be close ($\frac{1}{33.9982}$) to $\frac{1}{34}$ as large as the full-rank probability, presently conjectured to be $\frac{27}{1000}$.
(The rank-4 two-rebit HS separability probability has been {\it proven} by Lovas and Andai to equal $\frac{29}{64}=\frac{29}{2^6}$ \cite{lovas2017invariance}.)

\section{Analyses}
\subsection{Rebit-retrit analysis}
\begin{figure}
    \centering
    \includegraphics{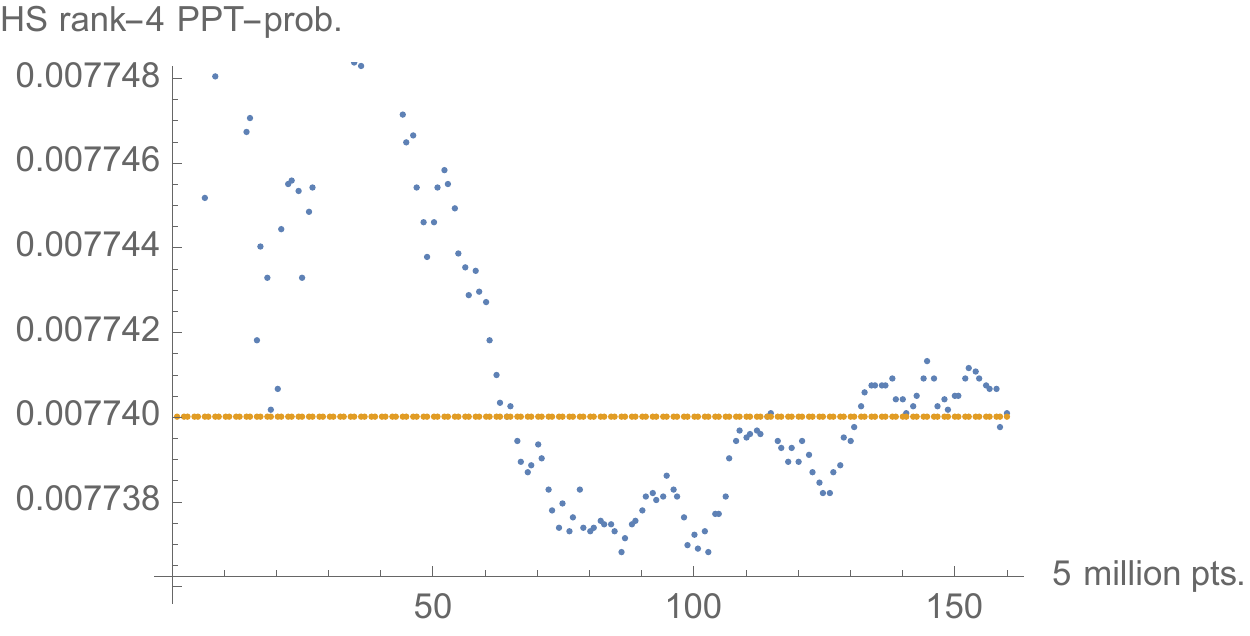}
    \caption{Estimates of rank-4 rebit-retrit Hilbert-Schmidt PPT/separability probability and conjectured value of $\frac{387}{50000} =\frac{3^2 \cdot 43}{2^4 \cdot 5^5}=0.00774$.}
    \label{fig:RebitRetritPlot}
\end{figure}

Implementing the indicated procedure for rank-4 rebit-retrit states, for 800 million Ginibre-matrix realizations, 6,192,047 were found separable for a sample probability of .00774006--suggestive of an exact value of $\frac{387}{50000} =\frac{3^2 \cdot 43}{2^4 \cdot 5^5}=.00774$ (Fig.~\ref{fig:RebitRetritPlot}). 
Subject to such a conjecture and the indicated $\frac{860}{6561} =\frac{2^2 \cdot 5 \cdot 43}{3^8}$ full-rank one, the ratio of the rank-4 to rank-6 probabilities would be 
$\frac{59049}{1000000}=\frac{3^{10}}{2^6 \cdot 5^6} =(\frac{243}{1000})^2 \approx 0.059049$, with the common factor 43 interestingly cancelling. For ranks of three and less, the 2009 theorem of Ruskai and Werner \cite{ruskai2009bipartite} informs us that the associated separability probabilities are zero.
\subsection{Qubit-qutrit analysis}
However, this new--to us, intriguing--rebit-retrit conjecture of $\frac{59049}{1000000}=\frac{3^{10}}{2^6 \cdot 5^6}$, seemed somewhat different (perhaps more ``elegant") in nature than--at this point in time--its apparent qubit-qutrit counterpart, which would involve dividing $\frac{27}{1000}$ by $\frac{1}{34}$,
yielding $\frac{27}{34000}=\frac{3^3}{2^4 \cdot 5^3 \cdot 17} \approx 0.000794118$. Since the ``34" stemmed from an estimate of 33.9982 reported in our ``long ago" 2005 study \cite{slater2005qubit}--relying upon quasi-Monte Carlo (Tezuka-Faure) numerical integration--we decided to re-examine it employing the new, detailed-above  Wishart-Laguerre-based methodology of K. {\.Z}yczkowski and the group of A. Khvedelidze, I. Rogojin and V. Abgaryan. 
 
\begin{figure}
    \centering
    \includegraphics{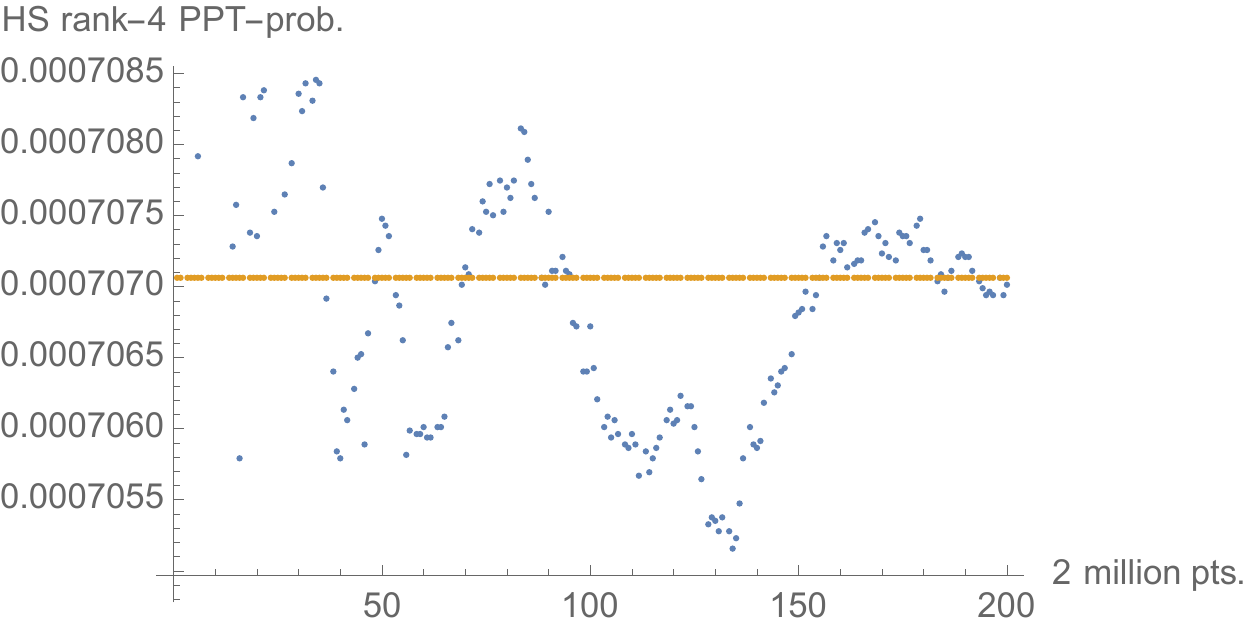}
    \caption{Estimates of rank-4 qubit-qutrit Hilbert-Schmidt PPT/separability probability and conjectured value of $\frac{7}{9900} =\frac{7}{2^2 \cdot 3^2 \cdot 5^2 \cdot 11}=0.000707071$.}
    \label{fig:QubitQutritPlot}
\end{figure}

Then, employing for hundred million $4 \times 8$ complex-entry Ginibre matrices, we obtained an estimate of 0.000707020, of similar magnitude, but still markedly different from the indicated 0.000794118 (Fig.~\ref{fig:QubitQutritPlot}). (In the 2005 study, contrastingly, an Euler-angle parameterization of unitary matrices was employed. But it is not now quite clear there, in what manner the parameterization was adopted to the rank-4 analysis.) This result is  suggestive of an exact value of $\frac{7}{9900} =\frac{7}{2^2 \cdot 3^2 \cdot 5^2 \cdot 11}=0.000707071$.
Subject to this conjecture and the indicated  $\frac{27}{1000}$ full-rank one, the ratio of the rank-4 to rank-6 probabilities would be 
$\frac{70}{2673}=\frac{2 \cdot 5 \cdot 7}{ 3^5  \cdot 11} \approx 0.0261878$. 
\subsection{Qubit-ququart analysis}
In \cite{slater2019numerical}, a Hilbert-Schmidt PPT-probability conjecture of $\frac{16}{12375} = \frac{2^4}{3^2 \cdot 5^2 \cdot 11} \approx 0.0012929$ was advanced for $2 \times 4$ qubit-ququart systems. In a further analysis of ours, for rank-6 such systems, based on 149 million Ginibre-matrix realizations, we obtained a PPT-probability estimate of 0.0000546242. Though we plan to extend this analysis, a tentative conjecture for this last value is $\frac{169}{3093750}= \frac{13^2}{2 \cdot 3^2 \cdot 5^6 \cdot 11} \approx 0.0000546263$ with the rank-6/rank-8 ratio, then, being $\frac{169}{4000} =\frac{13^2}{2^5 \cdot 5^3} \approx 
0.04225$.
\section{Concluding Remarks}
In the course of the research reported above, we have, in particular, sought rational-valued Hilbert-Schmidt PPT/separability rebit-retrit and qubit-qutrit probability formulas. 
Certainly, we have no demonstration that this must, in fact, be the case. But in light of the proven nature of the two-rebit probability ($\frac{29}{64}$) \cite{lovas2017invariance}, and the strong evidence for the two-qubit ($\frac{8}{33})$, two-quater[nionic]bit ($\frac{26}{323}$),\ldots,counterparts  
\cite{slater2018master}, this seems a direction worth pursuing--especially in light of the elegant nature of the formulas so far found (not to mention also the ``half-theorem" of Szarek, Bengtsson and {\.Z}yczkowski \cite{szarek2006structure}). Also, in terms of the Hilbert-Schmidt measure, the two-qubit separability probability is equally divided between those states for which $|\rho| > |\rho^{PT}|$ and those for which $|\rho^{PT}| > |\rho|$ \cite{slater2018formulas}.

Needless to say, it would seem, the intrinsic high-dimensionality (twenty and thirty-five) of the problems under consideration above, and related ones, presents formidable challenges to exact, symbolic analyses, as contrasted with the numerical approach adopted here.

\begin{acknowledgements}
This research was supported by the National Science Foundation under Grant No. NSF PHY-1748958. I thank 
K. {\.Z}yczkowski and the research group of A. Khvedelidze, I. Rogojin and V. Abgaryan for their several communications.
\end{acknowledgements}

\bibliography{main}

\end{document}